\documentclass[aps,prb,twocolumn,superscriptaddress,showpacs,floatfix,altaffilletter]{revtex4-1}

\usepackage{graphicx}
\usepackage{bm}
\usepackage{amsmath}
\usepackage{sansmath}
\usepackage{amsfonts}
\usepackage{amssymb}
\usepackage{xcolor}
\usepackage{xspace}
\newcommand{\alio}{$\alpha$-Li$_2$IrO$_3$\xspace}
\newcommand{\blio}{$\beta$-Li$_2$IrO$_3$\xspace}
\newcommand{\glio}{$\gamma$-Li$_2$IrO$_3$\xspace}

\begin{document}
\title{Incommensurate Counterrotating Magnetic Order Stabilized by \\
Kitaev Interactions in the Layered Honeycomb
$\bm{\alpha}$-Li$_2$IrO$_3$}

\author{S.\,C.~Williams}
\affiliation{Clarendon Laboratory, University of Oxford Physics
Department, Parks Road, Oxford, OX1 3PU, United Kingdom}
\author{R.\,D.~Johnson}
\affiliation{Clarendon Laboratory, University of Oxford Physics
Department, Parks Road, Oxford, OX1 3PU, United Kingdom}
\affiliation{ISIS Facility, Rutherford Appleton Laboratory-STFC,
Chilton, Didcot, OX11 0QX, United Kingdom}
\author{F.~Freund}
\affiliation{EP VI, Center for Electronic Correlations and
Magnetism, Augsburg University, D-86159 Augsburg, Germany}
\author{Sungkyun\ Choi}
\altaffiliation[Current address: ] {Max Planck Institute for Solid
State Research, Heisenbergstra{\ss}e 1, 70569 Stuttgart, Germany}
\affiliation{Clarendon Laboratory, University of Oxford Physics
Department, Parks Road, Oxford, OX1 3PU, United Kingdom}
\author{A.~Jesche}
\affiliation{EP VI, Center for Electronic Correlations and
Magnetism, Augsburg University, D-86159 Augsburg, Germany}
\author{I.~Kimchi}
\affiliation{Department of Physics, Massachusetts Institute of
Technology, 77 Massachusetts Ave., Cambridge, MA 02139}
\author{S.~Manni}
\altaffiliation[Current address: ] {Ames Laboratory, Department of
Physics and Astronomy, Iowa State University, Ames, Iowa-50010}
\affiliation{EP VI, Center for Electronic Correlations and
Magnetism, Augsburg University, D-86159 Augsburg, Germany}
\author{A.~Bombardi}
\affiliation{Diamond Light Source Ltd., Harwell Science and
Innovation Campus, OX11 0DE, United Kingdom}
\author{P.~Manuel}
\affiliation{ISIS Facility, Rutherford Appleton Laboratory-STFC,
Chilton, Didcot, OX11 0QX, United Kingdom}
\author{P.~Gegenwart}
\affiliation{EP VI, Center for Electronic Correlations and
Magnetism, Augsburg University, D-86159 Augsburg, Germany}
\author{R.~Coldea}
\affiliation{Clarendon Laboratory, University of Oxford Physics
Department, Parks Road, Oxford, OX1 3PU, United Kingdom}

% 75.25.-j  Spin arrangements in magnetically ordered materials
% (including neutron and spin-polarized electron studies, synchrotron-source x-ray scattering, etc.
% 75.10.Jm  Quantized spin models, including quantum spin frustration

\begin{abstract}
The layered honeycomb magnet $\alpha$-Li$_2$IrO$_3$ has been
theoretically proposed as a candidate to display novel magnetic
behaviour associated with Kitaev interactions between spin-orbit
entangled $j_{\rm eff}=1/2$ magnetic moments on a honeycomb
lattice. Here we report single crystal magnetic resonant x-ray
diffraction combined with powder magnetic neutron diffraction to
reveal an incommensurate magnetic order in the honeycomb layers
with Ir magnetic moments counter-rotating on nearest-neighbor
sites. This type of magnetic structure has not been reported
experimentally before in honeycomb magnets and cannot be explained
by a spin Hamiltonian with dominant isotropic (Heisenberg)
couplings. The magnetic structure shares many key features with
the magnetic order in the structural polytypes $\beta$- and \glio,
understood theoretically to be stabilized by dominant Kitaev
interactions between Ir moments located on the vertices of
three-dimensional hyperhoneycomb and stripyhoneycomb lattices,
respectively. Based on this analogy and a theoretical soft-spin
analysis of magnetic ground states for candidate spin
Hamiltonians, we propose that Kitaev interactions also dominate in
$\alpha$-Li$_2$IrO$_3$, indicative of universal Kitaev physics
across all three members of the harmonic honeycomb family of
Li$_2$IrO$_3$ polytypes.
\end{abstract} \maketitle

%% Introduction
\section{Introduction}
\label{sec:intro} Magnetic materials in the strong spin-orbit
regime are attracting much interest as candidates to display novel
magnetic states stabilized by frustration effects from
bond-dependent anisotropic interactions.\cite{review_iridates} One
of the most theoretically studied Hamiltonians with bond-dependent
interactions is the Kitaev model on the honeycomb lattice, where
all bonds carry an Ising exchange, but the three bonds meeting at
each lattice site have reciprocally-orthogonal Ising axes (along
cubic $\mathsf{x},\mathsf{y}$ and $\mathsf{z}$ directions). This
leads to strong frustration effects that stabilize an
exactly-solvable quantum spin liquid ground state,\cite{kitaev}
with unconventional forms of magnetic order predicted to occur
when additional magnetic interactions perturb the pure Kitaev
limit.\cite{chaloupka,rau,rachel,Chaloupka15} $A_2$IrO$_3$
materials ($A$=Na, Li) with three-fold coordinated, edge-sharing
IrO$_6$ octahedra have been proposed\cite{jackeli,chaloupka} as
prime candidates to realize such physics as (i) the combination of
strong spin-orbit coupling and the near-cubic crystal field
stabilize a $j_{\rm eff}=1/2$ spin-orbit entangled magnetic moment
at the Ir site, (ii) for edge-sharing bonding geometry
superexchange between neighboring Ir moments is expected to be (to
leading order) of Ising form, coupling only the moment components
perpendicular to the plane of the Ir-O$_2$-Ir square plane of the
bond, and (iii) the three bonds emerging out of each Ir lattice
site have near-orthogonal Ir-O$_2$-Ir planes. These are key
ingredients for frustrated bond-dependent, anisotropic
interactions.

The first material to be explored in search of Kitaev physics was
Na$_2$IrO$_3$, which has a layered crystal structure where
edge-sharing IrO$_6$ octahedra form a honeycomb arrangement. The
Ir moments order magnetically\cite{singh} at low temperature in a
zigzag magnetic structure\cite{liu,choi,ye} (ferromagnetic zigzag
chains ordered antiferromagnetically in the honeycomb plane),
which was proposed to be stabilized by many competing
interactions.\cite{Chaloupka15} Evidence for Kitaev couplings was
provided by the observation of a locking between the spin
fluctuations direction and wavevector.\cite{Chun15} Li$_2$IrO$_3$
can also be prepared in an iso-structural form ($\alpha$-phase,
Ref.~\onlinecite{omalley}) with Na$^+$ replaced by Li$^{+}$.
Furthermore, two other structural polytypes, $\beta$-
[Ref.~\onlinecite{beta_takagi}] and \glio
[Ref.~\onlinecite{modic}] have also been recently synthesized.
Both latter structures share the same building blocks of
three-fold coordinated, edge-sharing IrO$_6$ octahedra, but rather
than being arranged in honeycomb layers, now the IrO$_6$ octahedra
form three-dimensionally connected structures, called
hyperhoneycomb and stripyhoneycomb, respectively. All three
polytypes can be systematically understood as members of a
``harmonic honeycomb" structural series.\cite{modic} This
multitude of structural polytypes for Li$_2$IrO$_3$ is attributed
to the fact that Li$^+$ and Ir$^{4+}$ have rather comparable ionic
radii (Na$^+$ is a much larger ion, so only the layered honeycomb
structure appears to form). Both $\beta$- and \glio show
incommensurate magnetic structures with counter-rotating
moments,\cite{beta,gamma} understood theoretically to be
stabilized by dominant Kitaev interactions and additional small
terms;\cite{gamma, ybkim1, ybkim2} surprisingly, the $\beta$ and
$\gamma$ magnetic structures are so similar that they can be
considered as ``equivalent",\cite{beta} leading to proposals of
universality of the magnetism in the family of harmonic honeycomb
iridates.\cite{universality}

Motivated by those ideas we have performed detailed experimental
studies of the magnetic order in the layered polytype \alio, for
which early susceptibility and specific heat measurements in
powder samples\cite{singh-manni} have indicated magnetic
long-range ordering below $\simeq$15\ K. No experimental studies
of the magnetic structure have been reported so far, however many
theoretical proposals have been put forward for rather exotic
magnetic
structures.\cite{rachel,rau,vandenBrink,universality,Chaloupka15}
On the honeycomb lattice many distinct types of magnetic orders
are symmetry allowed, especially for the case of an incommensurate
propagation vector, so a complete experimental magnetic structure
solution is required in order to provide vital constraints for
candidate theoretical models. Using a novel sample synthesis
method not applied to iridates before, we have recently obtained
phase-pure, single crystals of \alio and here we report magnetic
resonant x-ray diffraction (MRXD) measurements on those crystals,
combined with magnetic powder neutron diffraction measurements and
symmetry analysis to determine a complete magnetic structure
solution. We find an incommensurate magnetic order in the
honeycomb layers with counter-rotating Ir moments on every nearest
neighbor bond. We complement the experimental results with a
theoretical soft-spin analysis\cite{universality} and propose a
minimal nearest-neighbor spin Hamiltonian with dominant Kitaev
interactions and additional small terms, which naturally explains
the stability of the observed incommensurate structure and the
many common features with the magnetic structures in the $\beta$
and $\gamma$ polytypes. Our results emphasize that Kitaev
interactions between spin-orbit entangled $j_{\rm eff}=1/2$
Ir$^{4+}$ magnetic moments lead to universal magnetism in all
three members of the harmonic honeycomb Li$_2$IrO$_3$ polytypes.

%%%%%%%%%%%%%%%%%%%%%%%%%%%%%%%%%%%%%%%%%%%%%%%%%%%%%%%%%%%%%%%%
\begin{figure}[!tbhp]
\includegraphics[width=0.5\textwidth]{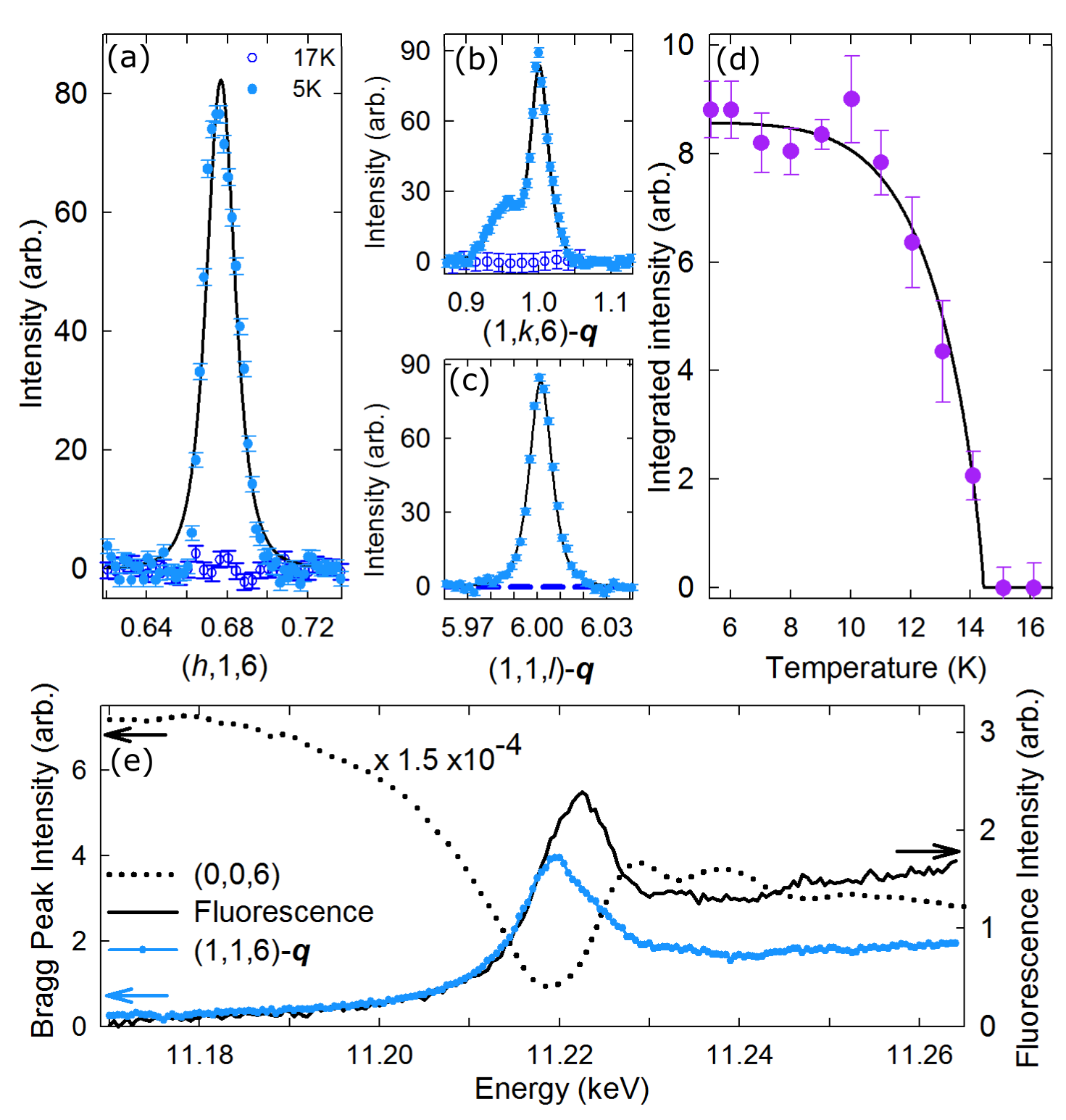}
\caption[]{(color online) Magnetic peak at $(1,1,6)$$-$${\bm q}$.
a-c) Scans along three different reciprocal space directions
(filled/open symbols are at base temperature/above $T_{\rm N}$).
Solid lines are fits to a Lorentzian-squared shape (panel (b)
shows a side shoulder attributed to the finite sample mosaic). d)
Temperature-dependence of the integrated peak intensity (solid
line is guide to the eye). e) Energy scan through the magnetic
peak (thick blue solid symbols) showing a large resonant
enhancement with a maximum at the onset edge of the fluorescence
signal from the sample (black solid line, scaled). In contrast,
the same energy scan through a structural peak (dotted line) shows
minimum intensity near resonance (due to increased x-ray
absorption). Data points in all panels are shown with an estimate
of the incoherent background subtracted off.}
\label{fig:braggpeak}
\end{figure}
%%%%%%%%%%%%%%%%%%%%%%%%%%%%%%%%%%%%%%%%%%%%%%%%%%%%%%%%%%%%%%%%%

%%%%%%%%%%%%%%%%%%%%%%%%%%%%%%%%%%%%%%%%%%%%%%%%%%%%%%%%%%%%%%%%
\begin{figure}[!tbhp]
\includegraphics[width=0.5\textwidth]{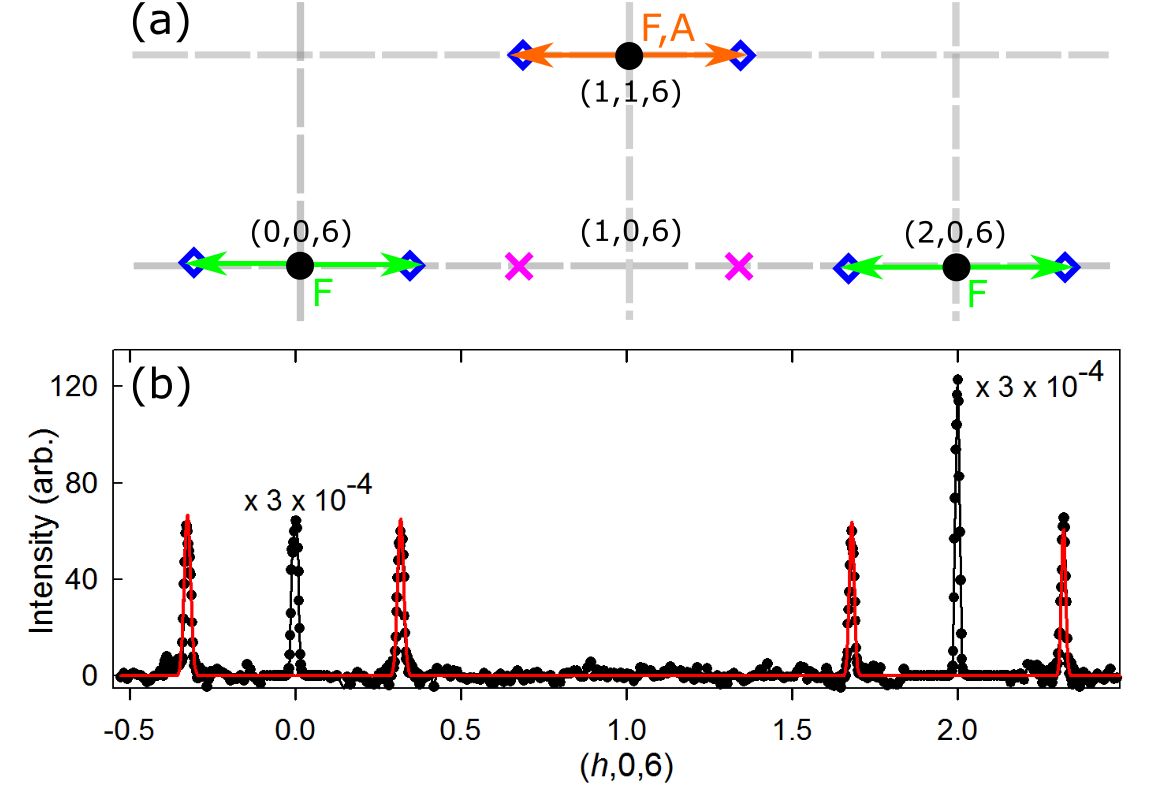}
\caption[]{(color online) a) Schematic diagram of the $(hk6)$
reciprocal plane with filled circles, diamonds and magenta crosses
indicating positions of structural peaks, measured magnetic peaks
and the absence of peaks, respectively. Lattice points are also
labelled by the magnetic basis vectors that have finite structure
factor for magnetic peaks at satellite $\pm{\bm q}$ positions. b)
Scan along the $(h,0,6)$ direction observing structural peaks at
integer $h=0,2$ (intensity scaled by $3\times 10^{-4}$ for
clarity), and magnetic peaks at satellite positions $h = 0{\pm}q,
2 \pm q$. Solid (red) line is the calculated magnetic scattering
intensity\cite{magnetix} for the magnetic structure model depicted
in Fig.~\ref{fig:magstruct}. Data points are raw counts with an
estimate of the incoherent background subtracted off.}
\label{fig:hscan}
\end{figure}
%%%%%%%%%%%%%%%%%%%%%%%%%%%%%%%%%%%%%%%%%%%%%%%%%%%%%%%%%%%%%%%%%

The paper is organized as follows: Sec.~\ref{sec:MRXD} presents
the single-crystal MRXD measurements, which observe magnetic
diffraction peaks with an incommensurate propagation vector ${\bm
q}=(0.32(1),0,0)$. The observed diffraction pattern is analyzed in
terms of magnetic basis vectors, and their polarization and
relative phase are determined from the azimuth dependence of the
diffraction intensities in Sec.~\ref{sec:basisvectors}. The
absolute value of the ordered magnetic moment is extracted from
neutron powder diffraction data in Sec.~\ref{sec:npd}. The
obtained magnetic structure is presented in
Sec.~\ref{sec:magstructure} and similarities with the magnetic
structures in the $\beta$- and $\gamma$-polytypes are discussed in
Sec.~\ref{sec:discussion}. Finally, conclusions are summarized in
Sec.~\ref{sec:conclusions}. The Appendices contain (A) technical
details of the magnetic symmetry analysis and the decomposition of
the magnetic structure in terms of its Fourier components, (B)
description of the crystal and magnetic structure of \alio in
terms of the orthorhombic axes common to the $\beta$ and $\gamma$
polytypes, (C) derivation of the direct link between the
counter-rotation of magnetic moments and the anti-phase behavior
of the MRXD intensity at $\pm{\bm q}$ magnetic satellites, and (D)
a theoretical analysis of the minimal model Hamiltonian that could
stabilize the observed magnetic structure in \alio.

%% Method
\section{Magnetic Resonant X-ray Diffraction}
\subsection{Experimental results}
\label{sec:MRXD} MRXD experiments were performed using the I16
beamline at Diamond with photon energies near the L$_3$ edge of
Ir. The sample was a single crystal of \alio (maximum dimension
$\sim$200~$\mu$m, the crystal synthesis and characterization is
described elsewhere\cite{friedrich}), which was placed with the
(001) axis approximately surface normal onto on a Si (111) plate,
and cooled using a closed-cycle refrigerator with a Be dome. With
the x-ray energy tuned to resonance at 11.217\, keV, and the
sample temperature set to $\approx5$ K, diffraction peaks were
observed at satellite positions $\bm{\tau} \pm \bm{q}$ of allowed
structural reflections, $\bm{\tau}=(h,k,l)$ with $h+k=$ even, and
with the propagation vector\cite{error} ${\bm q}=(0.32(1),0,0)$.
Throughout we label wave vectors in reciprocal lattice units of
the structural monoclinic unit cell with space group $C2/m$ (for
more details see Appendix A). A representative scan is shown in
Fig.~\ref{fig:braggpeak}a) (solid circles). Also shown are data
points collected at high temperature (17 K, open circles), which
illustrate that this diffraction signal is only present at low
temperatures. The temperature dependence of the integrated peak
intensity is shown in Fig.~\ref{fig:braggpeak}d), and was found to
have a typical order-parameter behavior with an onset temperature
$T_{\rm N}=14.4(2)$\, K, which essentially coincides with the
transition temperature to magnetic order inferred from earlier
specific heat and susceptibility measurements on powder
samples.\cite{singh-manni} We therefore attribute the satellite
peaks to x-ray diffraction from the periodic magnetic order of Ir
moments. The satellite peaks were as sharp as structural peaks in
scans along all three reciprocal space directions (representative
scans shown in Fig.~\ref{fig:braggpeak}a-c), indicating coherent,
3-dimensional long-range magnetic order. The magnetic origin of
the satellite reflections is further confirmed by the intensity
dependence on the x-ray energy. Fig.~\ref{fig:braggpeak}e)(blue
solid symbols) shows that the peak intensity has a large resonant
enhancement, as characteristic of magnetic x-ray diffraction. The
empirically observed x-ray resonance energy is similar to values
found in other iridates \cite{liu,beta,gamma} and agrees well with
the edge of the measured fluorescence signal from the sample
(black solid line in Fig.~\ref{fig:braggpeak}e).

We note that the observed propagation vector ${\bm q}$ is close to
the commensurate wavevector $(1/3,0,0)$, which corresponds to an
exact tripling of the unit cell along $a$, however, this
commensurate wavevector is not a special high symmetry point in
the Brillouin zone of the structural $C2/m$ space group, but has
the same symmetry as any general point in the $(h0l)$ plane. In
the following analysis of the magnetic structure we therefore
treat ${\bm q}$ as a general incommensurate wavevector. The fact
that ${\bm q}$ has no component along ${\bm c}^*$ has a natural
physical interpretation: adjacent honeycomb layers are stacked
ferromagnetically along ${\bm c}$.

%% Analysis
\subsection{Magnetic basis vectors}
\label{sec:basisvectors} Systematic surveys in reciprocal space
revealed that satellite peaks occurred only around structural
Bragg peaks. For example Fig.~\ref{fig:hscan}b) shows a scan along
the ($h,0,6$) direction where the red solid line highlights the
observed magnetic peaks at $h=0\pm q $ and $2\pm q$, with no
magnetic signal at $h=1\pm q$ (magenta crosses in
Fig.~\ref{fig:hscan}a); several azimuth values were tested, not
shown). Therefore, the magnetic structure can be fully described
in terms of Fourier components of magnetic moments located in the
structural primitive cell. \alio has a monoclinic crystal
structure with space group $C2/m$ with two Ir atoms in the
primitive cell, labelled here as Ir1 at $(0,y,0)$ and Ir2 at
$(0,-y,0)$ with $y=0.3332$, where the atomic fractional
coordinates are given in the $C2/m$ cell.\cite{omalley,nominal}
For a propagation vector ${\bm q}=(q,0,0)$ symmetry
analysis\cite{basireps} in the $C2/m$ space group gives two
magnetic basis vectors with Fourier components at the two iridium
sites in-phase or in anti-phase, in short-hand notation labelled
$F$ and $A$, respectively. The structure factors for the two basis
vectors for a magnetic reflection at wavevector $\bm{Q}=(h,k,l)
\pm {\bm q}$ are
\begin{eqnarray}
{\cal S}^F & = & 2f_{C} ~\cos(2\pi k y) ,\nonumber\\
{\cal S}^A & = & 2f_{C} ~i\sin(2\pi k y), \label{eq:sf}
\end{eqnarray}
where the pre-factor $f_{C}=1+e^{i\pi (h+k)}$ arises from the
$C$-centering in the $ab$ plane. Using the approximation $y \simeq
1/3$ implies that $F$-basis vectors can contribute to magnetic
satellites of all structurally-allowed peaks ($h+k=$ even),
whereas $A$-basis vectors could contribute only to the subset of
those with $k \neq 3n$, $n$ integer. Below we use those selection
rules and the polarization dependence of the MRXD cross-section to
determine which basis vectors are present, their polarization, and
relative phase.

%%%%%%%%%%%%%%%%%%%%%%%%%%%%%%%%%%%%%%%%%%%%%%%%%%%%%%%%%%%%%%%%%%
\begin{figure}[!tbhp]
\includegraphics[width=0.46\textwidth]{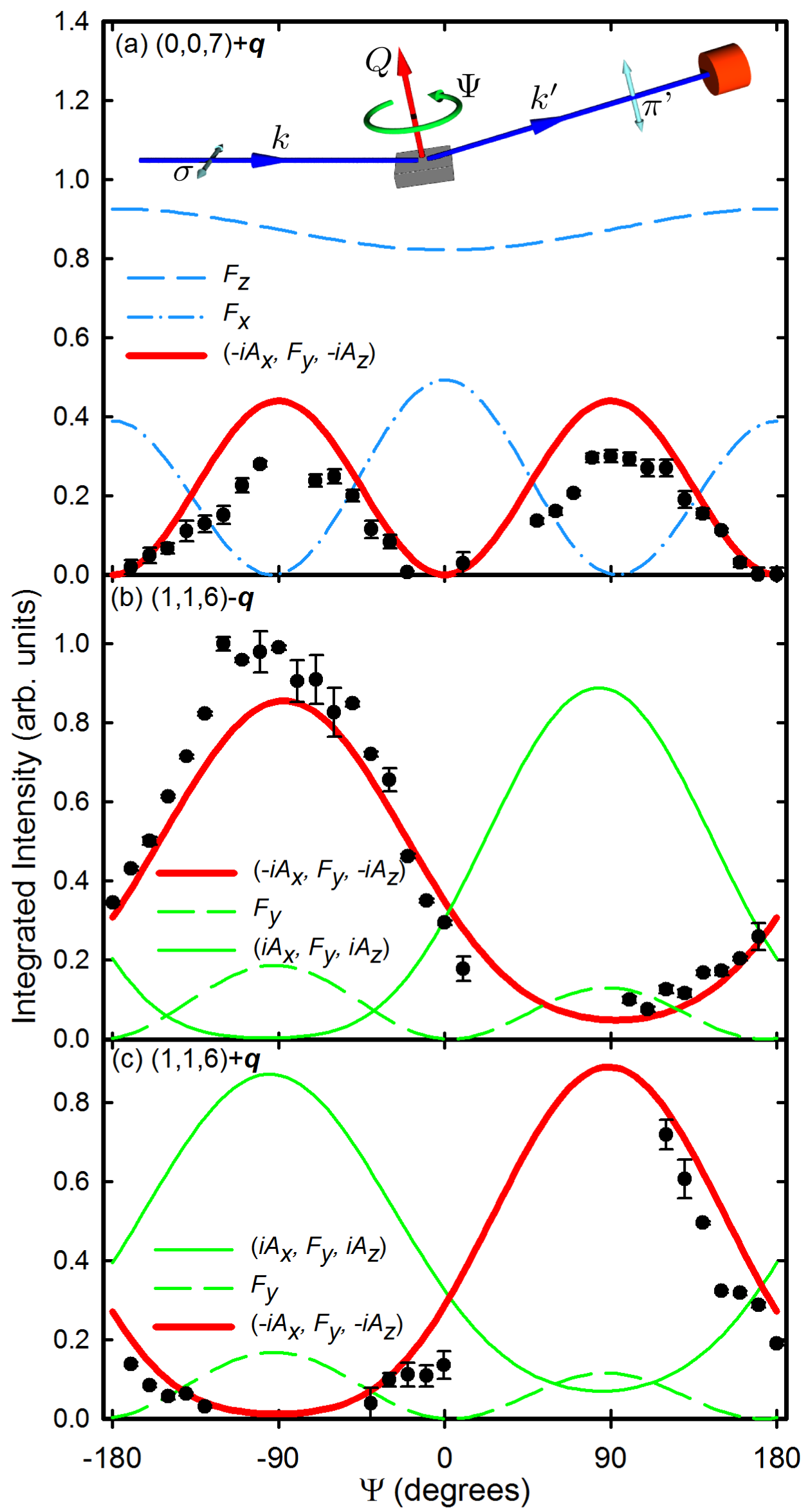}
\caption[]{(color online) Integrated intensity as a function of
azimuth for three magnetic Bragg peaks, a) pure-$F$, b) and c)
paired satellites of mixed-$FA$ character. Top diagram illustrates
the scattering geometry. Data points (filled circles) are
integrated peak intensities from rocking curve scans corrected for
absorption and Lorentz factor. Thick (red) lines show fits that
include all contributions to the MRXD structure factor
\cite{magnetix} for the magnetic structure model
$(-iA_x,F_y,-iA_z)$, depicted in Fig.~\ref{fig:magstruct}.
Blue/green curves illustrate that other phase combinations of
basis vectors are ruled out.} \label{fig:azimuths}
\end{figure}
%%%%%%%%%%%%%%%%%%%%%%%%%%%%%%%%%%%%%%%%%%%%%%%%%%%%%%%%%%%%%%%%%%

For a $\sigma$-polarized incident beam (electric field normal to
the scattering plane) only the projection of the magnetic moments
onto the scattered beam direction, $\bm{\hat{k'}}$, contributes to
the diffraction intensity.\cite{hill} By rotating the sample
around the scattering vector, ${\bm Q}=\bm{k'}-\bm{k}$, by the
azimuth angle, $\Psi$, [see diagram in Fig.~\ref{fig:azimuths}a)
inset] the projection of the magnetic moments onto $\bm{\hat{k'}}$
changes, giving a clear signature of the moment direction (in the
following we employ a convenient Cartesian set of axes
($x$,$y$,$z$) derived from the monoclinic axes, ${\bm x}
\parallel \bm{a}$, ${\bm y} \parallel {\bm b}$ and ${\bm z}
\parallel {\bm c}^*$, to describe magnetic moment directions). We have
measured the azimuth dependence for three magnetic peaks close to
the sample surface normal, such that the $\Psi$ rotation is almost
around (001). The origin, $\Psi=0$, is defined as the azimuth when
the (100) direction is in the scattering plane and pointing away
from the x-ray source. Fig.~\ref{fig:azimuths}a) shows the azimuth
scan for a pure-$F$ magnetic Bragg peak, $(0,0,7)+{\bm q}$ (${\cal
S}^A|_{k=0} = 0$). The intensity drops to essentially zero at
$\Psi=0$ and $\pm180^{\circ}$ and has maxima near $\pm90^{\circ}$,
uniquely identifying this signal as originating from diffraction
by $y$-moment components, \emph{i.e.} moments parallel to the
crystallographic $b$-axis (solid red line). Scattering from $x$-
and $z$-moment components (shown by dash-dotted and dashed lines,
respectively) have a qualitatively different behavior and can be
clearly ruled out. This analysis identifies the presence of a
basis vector component $F_y$ and the absence (within experimental
accuracy) of $F_x$ and $F_z$. Fig.~\ref{fig:azimuths}b) and c)
show the azimuth dependence of the intensity for the paired
magnetic satellites $(1,1,6)\mp{\bm q}$, where both $F$ and $A$
basis vectors can contribute. A pure $F_y$ basis vector (dashed
line) cannot explain the observed periodicity of the azimuth
dependence, and fails to predict the observed anti-phase behavior
of the intensity of the two satellites. The data is naturally
explained by adding an $A$ basis vector component to the magnetic
ground state with a comparable magnetic moment magnitude to the
$F_y$ component, polarized in the $xz$ plane, and with a $\pi/2$
phase difference. This basis vector combination, namely
$(-iA_x,F_y,-iA_z)$, was fit to the data as shown by thick red
lines in Fig.~\ref{fig:azimuths}a-c), which gives a good account
of the observed angular intensity dependence for all three azimuth
scans. All other basis vector combinations are ruled out
qualitatively by the data as illustrated by various (thin) lines
in the figures. The fit gives the relative magnetic moment
magnitudes as $M_x : M_y : M_z = 0.12(2):1: 0.74(4)$.

We note that the empirically determined basis vector combination,
$(-iA_x,F_y,-iA_z)$, corresponds to a single irreducible
representation, $\Gamma_1$, as listed in Table~\ref{tab:irr}. The
form of the magnetic structure is therefore fully consistent with
a continuous transition from paramagnetic to magnetic order below
$T_{\rm N}$.

The absolute magnitude of the ordered magnetic moments is
difficult to extract reliably from the MRXD data as it requires
accurate determination of scale factors between the magnetic and
structural peaks (the latter being $\sim10^4$ more intense, see
Fig.~\ref{fig:hscan}a). For this purpose we use neutron
diffraction where the structural and magnetic neutron scattering
factors are comparable, allowing one to reliably extract the
magnetic scattering intensity in absolute units.
%%%%%%%%%%%%%%%%%%%%%%%%%%%%%%%%%%%%%%%%%%%%%%%%%%%%%%%%%%%%%%%%%%%
\begin{table}[!btph] \caption{\label{tab:irr} Irreducible
representations and basis vectors for a magnetic structure with
propagation vector $\bm{q}$$=$$(q,0,0)$. The labels in brackets
correspond to the Miller and Love notation
convention.\cite{MillerandLove}}
\par
\begin{center}
\begin{tabular}{c|c}
Irreducible & Basis  \\
Representation & Vectors \\
\hline
$\Gamma_1$(B$_1$) & $A_x, F_y, A_z$\\
$\Gamma_2$(B$_2$) & $F_x, A_y, F_z$\\
\end{tabular}
\end{center}
\end{table}
%%%%%%%%%%%%%%%%%%%%%%%%%%%%%%%%%%%%%%%%%%%%%%%%%%%%%%%%%%%%%%%%%%%
%%%%%%%%%%%%%%%%%%%%%%%%%%%%%%%%%%%%%%%%%%%%%%%%%%%%%%%%%%%%%%%%%%%
\begin{figure}[!tbph]
\includegraphics[width=0.48\textwidth]{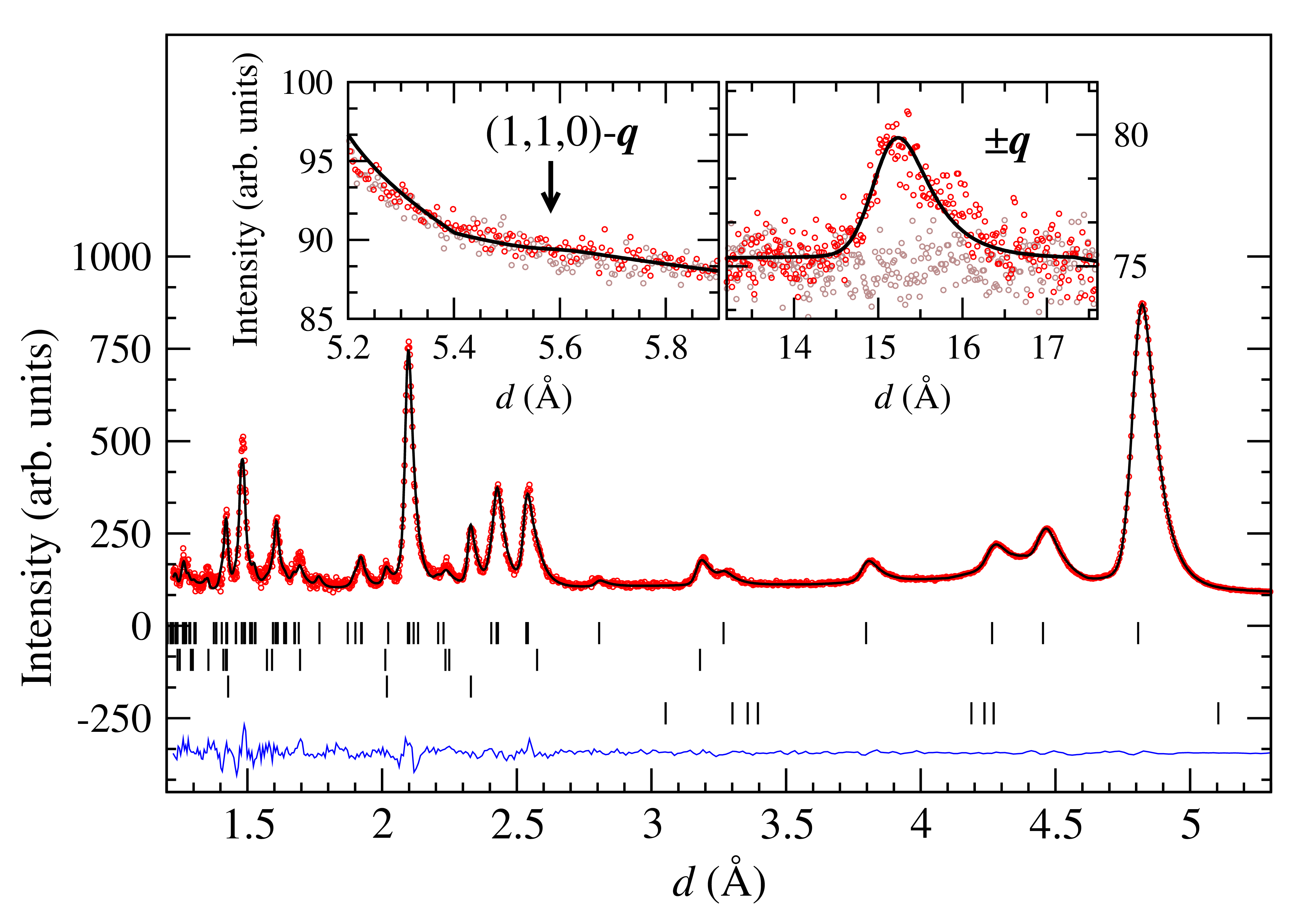}
\caption[]{(color online) Neutron powder diffraction at base
temperature (5.9\, K, red circles) and in the paramagnetic regime
(30\, K, brown circles) in the lowest-angle detector bank.
Positions of structural peaks, aluminium peaks (from the sample
sachet) and magnetic Bragg peaks are marked below the pattern in
the upper, middle and lower rows, respectively and the blue line
underneath represents the difference between data and fit. Insets:
zoom into the large $d$-spacing region showing the fundamental
magnetic peak indexed as $(000)\pm{\bm q}$ (right panel). In all
panels the solid black line shows the fit (using FullProf
\cite{fullprof}) to the structural and magnetic contributions as
discussed in the text.} \label{fig:npd}
\end{figure}
%%%%%%%%%%%%%%%%%%%%%%%%%%%%%%%%%%%%%%%%%%%%%%%%%%%%%%%%%%%%%%%%%%%

\section{Neutron Powder Diffraction}
\label{sec:npd} Neutron diffraction measurements were performed on
a 1.2~g powder sample of $\alpha$-Li$_2$IrO$_3$ (synthesized as
described in Ref.~\onlinecite{singh-manni}) using the
time-of-flight diffractometer WISH at ISIS. Powder
$\alpha$-Li$_2$IrO$_3$ is susceptible to absorb moisture when in
contact with air, which leads to a strong background signal due to
incoherent neutron scattering from the absorbed hydrogen. To
minimize this effect the sample was heated to a temperature of
110$^{\circ}$C under a continuously pumped vacuum (10$^{-5}$~bars)
for over 48~hours immediately prior to the neutron diffraction
experiments. The sample was placed in an aluminium sachet shaped
into an annular cylinder (to minimize neutron absorption) and
located inside a thin-walled vanadium can. Cooling was provided by
a closed cycle refrigerator and the neutron diffraction pattern
was collected at a selection of temperatures from base (5.9~K) to
paramagnetic (30~K).

Fig.~\ref{fig:npd} shows the measured diffraction pattern in the
lowest angle bank of detectors and the right inset shows a zoom of
the large $d$-spacing region where the fundamental magnetic peak
indexed as $(000)\pm{\bm q}$ was clearly observed. We fit
simultaneously three contributions to the diffraction data:
structural peaks of the sample, structural peaks of the aluminium
sachet containing the sample, and magnetic peaks of the sample.
The diffraction pattern did not allow for a full refinement of the
\alio crystal structure due to relative peak intensities being
affected by neutron absorption from iridium nuclei. We therefore
fixed the internal atomic positions to those reported by
room-temperature x-ray studies,\cite{omalley} and only refined the
lattice parameters and the atomic displacement parameters. This
strategy was found to be sufficient for scaling the magnetic
moment magnitude. The magnetic structure model deduced from the
resonant x-ray data in Sec.~\ref{sec:basisvectors} with the basis
vector combination ($-iA_x,F_y,-iA_z$), and fixed magnitude ratios
$M_x/M_y$ and $M_z/M_y$, was fitted to the data with only the
magnetic moment amplitude $M_y$ free to vary. The overall fit is
plotted as a solid black line in Fig.~\ref{fig:npd} and shows
excellent agreement with the data, both for the structural pattern
(main panel), as well as for the magnetic pattern (insets). In
particular, we note that the model accounts very well for the
observed strong intensity of the fundamental magnetic peak (right
inset) and essentially zero measurable intensity at the nominal
position of the second allowed magnetic peak (left inset). The
obtained ordered magnetic moment magnitude is 0.40(5)$\mu_{\rm B}$
when aligned along the $b$-axis, which is a lower bound owing to
attenuation of the diffraction peak intensity by Ir absorption.
The propagation vector was also fitted and found to be
$\bm{q}$$=$$(0.319(5),0,0)$, consistent with the value deduced
from single-crystal x-ray measurements.

%% Magnetic Structure
\section{Magnetic structure} \label{sec:magstructure} 
\begin{figure}[!tbhp]
\includegraphics[width=0.45\textwidth]{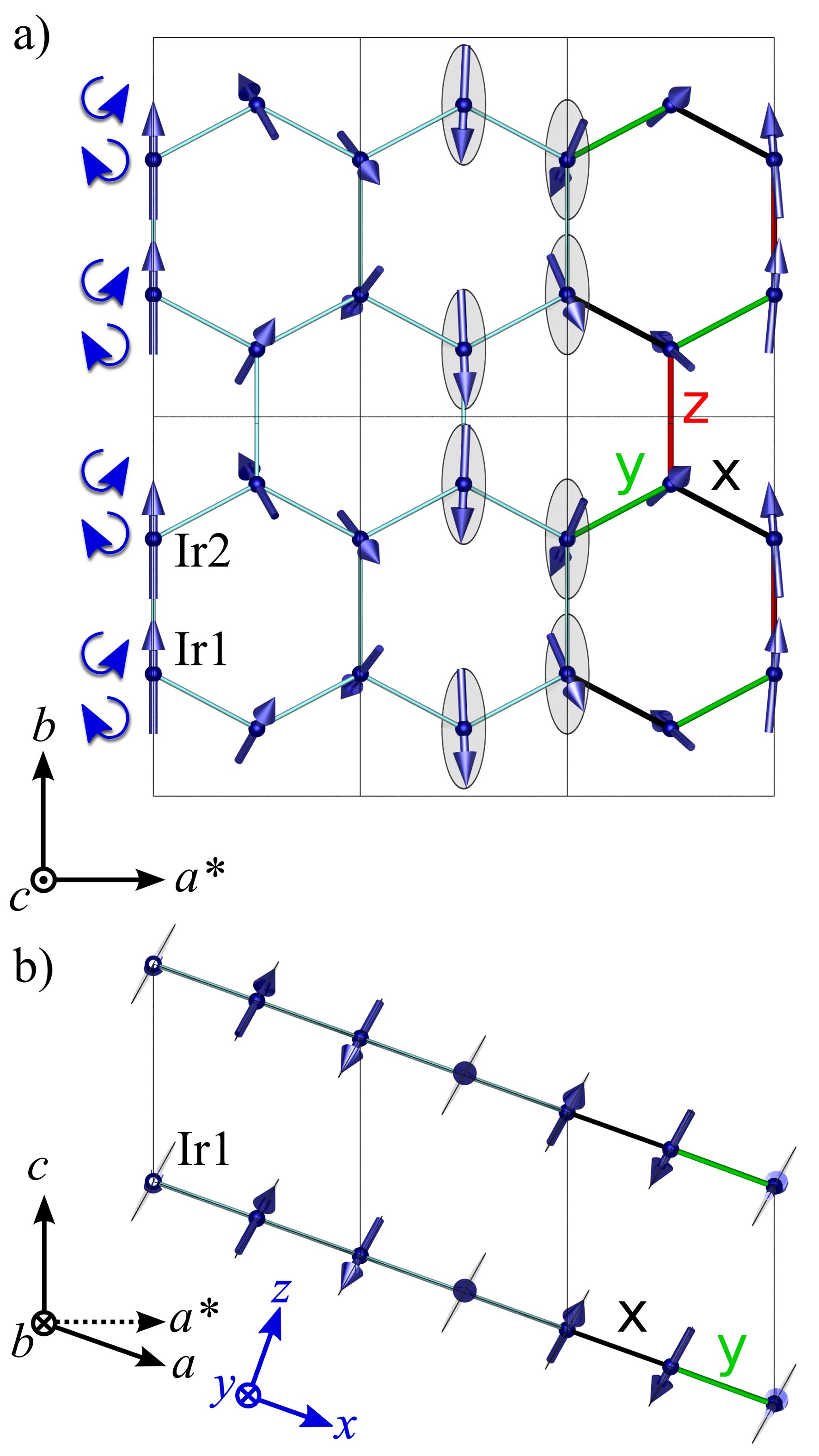}
\caption[]{(color online) a) Magnetic structure in a honeycomb
layer of \alio viewed along the monoclinic $\bm{c}$-axis. Three
unit cells are shown horizontally (along the propagation
direction) and two vertically, with unit cell edges indicated by
thin gray rectangles. The global phase of the moment rotation was
chosen such as to have the magnetic moments at the origin pointing
straight up along the $b$-axis. Left curly arrows illustrate
counter-rotation of the magnetic moments between consecutive sites
along $\bm{b}$. In unit cell 2 the light shaded ellipses show the
envelopes of the moment rotation. In unit cell 3 the color coding
of the bonds indicates the anisotropy axes of Kitaev exchange
(black, green, red for $\mathsf{x}$, $\mathsf{y}$, $\mathsf{z}$,
respectively). b) Projection of the magnetic structure onto the
${\bm a}{\bm c}$ plane showing ferromagnetic order between
adjacent layers stacked along $c$. The magnetic propagation vector
is along the horizontal direction ($\bm{a^*}$, dashed arrow). Thin
gray lines at each site give the projection of the elliptical
envelopes of moment rotation. The Cartesian axes ($x,y,z$) used to
describe the magnetic moment components are shown in blue at the
bottom of the figure.} \label{fig:magstruct}
\end{figure}

Having
determined the magnetic basis vectors, their amplitudes, and
relative phases, the magnetic structure in real space is obtained
via Fourier transformation as detailed in Appendix A,
eq.~(\ref{eq:Mr}). The resulting magnetic structure for one
honeycomb layer is plotted in Fig.~\ref{fig:magstruct}a). We show
the projection along the monoclinic $c$-axis to better visualize
the elliptical envelopes described by the rotation of the magnetic
moments when displaced along the (horizontal) propagation
direction. The elliptical envelopes have aspect ratio near 3:4
with the long axis along $b$, and they are oriented in a plane
that is almost normal to the $ab$ honeycomb layer (the precise
orientation w.r.t. the honeycomb layer is obtained by rotation
around the $b$-axis by an angle ${\rm
tan}^{-1}\frac{M_z}{M_x}=80.8\pm1.5^{\circ}$). This tilt is
illustrated in Fig.~\ref{fig:magstruct}b), which shows the
projection of the magnetic structure onto the $ac$ plane. An
important feature of the magnetic structure highlighted in
Fig.~\ref{fig:magstruct}a) is that nearest-neighbor sites in the
honeycomb lattice counter-rotate, this is true both for
nearest-neighbors of the zigzag chains along $a$, as well as for
vertically-connected sites along $b$, see left curly arrows in the
figure. The counter-rotation is a direct consequence of the basis
vector combination $F_y$ with $iA_{x,z}$, which means that for Ir1
and Ir2 sites in the same primitive cell the $y$ moment components
are parallel, whereas their perpendicular components in the $xz$
plane are antiparallel, leading to counter-rotation of the moments
on the two Ir sublattices. In Appendix \ref{app:interference} we
show that the counter-rotation of moments has a characteristic
signature in the MRXD intensity via an interference scattering
term that alternates in sign between $\pm{\bm q}$ satellites of
the same reciprocal lattice point, which leads to an anti-phase
behavior of the intensity in azimuth scans. For a given azimuth
value, at one satellite the interference term is added and at the
other satellite it is subtracted, so when one magnetic satellite
is strong the other is weak, and vice versa. This effect is
directly observed in our experiments at the paired satellites
$(116)\mp{\bm q}$ in Fig~\ref{fig:azimuths}b-c), which reveal a
pronounced anti-phase behavior of the intensity at the two
positions; this qualitative behavior of the intensity cannot be
explained by any other type of magnetic structure (for more
details see Appendix \ref{app:interference}).

%%%%%%%%%%%%%%%%%%%%%%%%%%%%%%%%%%%%%%%%%%%%%%%%%%%%%%%%%%%%%%%%%%
%%%%%%%%%%%%%%%%%%%%%%%%%%%%%%%%%%%%%%%%%%%%%%%%%%%%%%%%%%%%%%%%%%

\begin{figure}[!tbhp]
\includegraphics[width=0.48\textwidth]{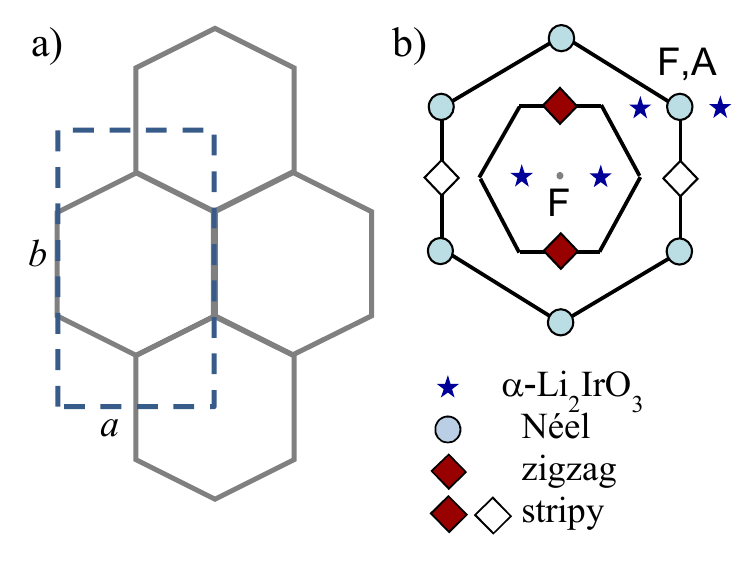}
\caption[]{(color online) a) Honeycomb lattice showing the $a
\times b$ unit cell (dashed rectangle). b) Reciprocal space
diagram of the honeycomb lattice showing the position of the
magnetic Bragg peaks (blue stars) corresponding to the
incommensurate magnetic order in \alio. Labels $F$ and $A$ next to
zone center positions $\bm{\tau}$ indicate the character of the
magnetic basis vectors that can contribute to the intensity of the
corresponding magnetic Bragg peaks at $\bm{\tau}\pm\bm{q}$. The
inner solid line hexagon is the 1st Brillouin zone of the
honeycomb lattice, and the other symbols are magnetic Bragg peak
positions for other types of magnetic structures such as N\'{e}el,
``zigzag'' with spins ferromagnetically aligned on the zigzag
bonds and antialigned along the vertical bonds, and ``stripy''
with spins ferromagnetically aligned along the vertical bonds and
antialigned along the zigzag bonds.}\label{fig:bz}
\end{figure}

For a direct comparison between the observed magnetic structure
and theoretical models for a two-dimensional honeycomb lattice we
show in Fig.~\ref{fig:bz}b) a diagram of the reciprocal space of
such a two-dimensional honeycomb, where the blue stars indicate
the location of the empirically determined magnetic Bragg peaks of
a single honeycomb layer of \alio. In this case the magnetic
propagation vector has components ($q,0$) with reference to a
rectangular $a \times b$ unit cell (dashed rectangle in
Fig.~\ref{fig:bz}a) of the honeycomb lattice.

%%%%%%%%%%%%%%%%%%%%%%%%%%%%%%%%%%%%%%%%%%%%%%%%%%%%%%%%%%%%%%%%%%%
%%%%%%%%%%%%%%%%%%%%%%%%%%%%%%%%%%%%%%%%%%%%%%%%%%%%%%%%%%%%%%%%%%%

%%%%%%%%%%%%%%%%%%%%%%%%%%%%%%%%%%%%%%%%%%%%%%%%%%%%%%%%%%%%%%%%%%%
\begin{figure}[!tbhp]
\includegraphics[width=0.45\textwidth]{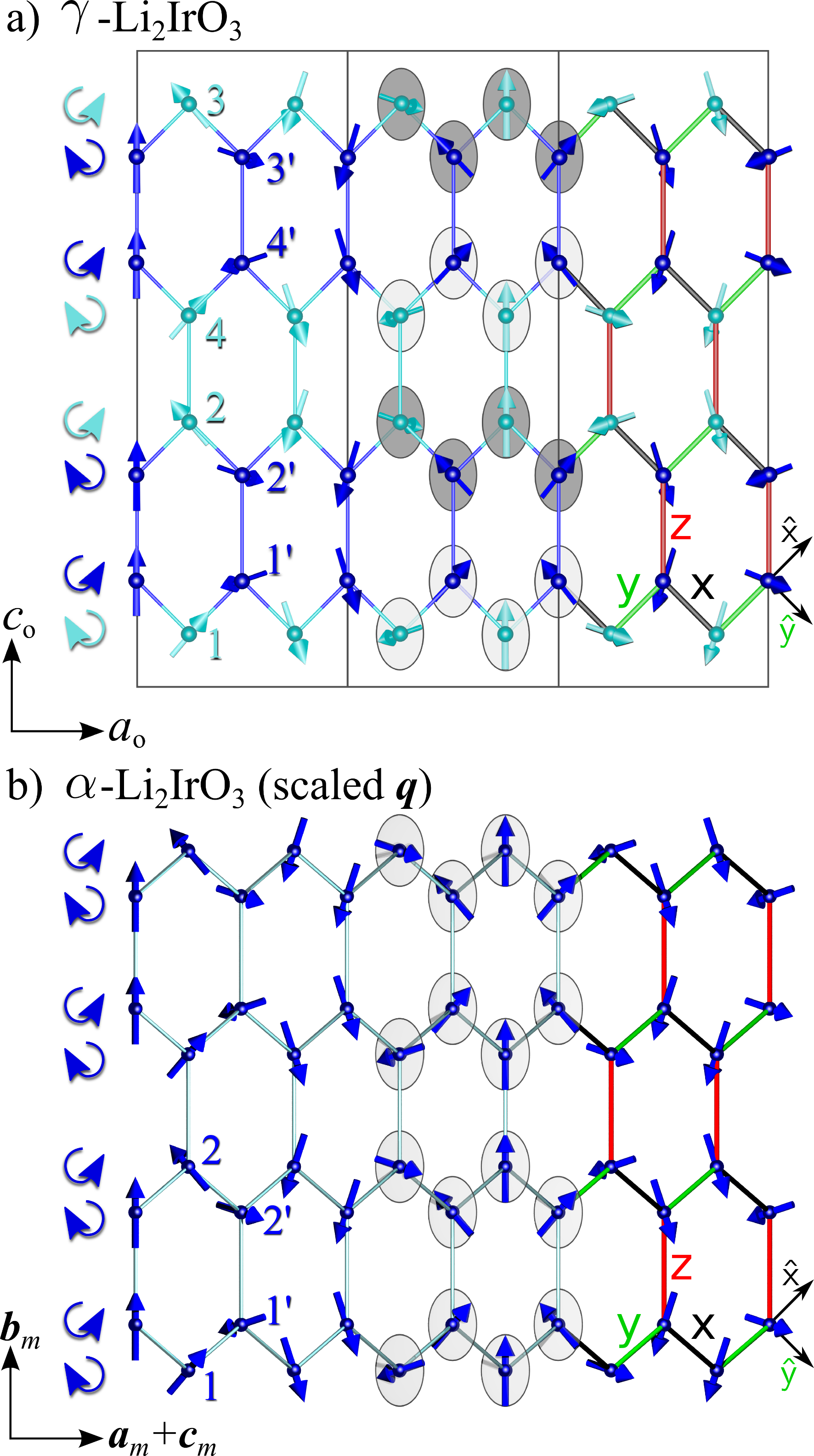}
\caption[]{(color online) a) Magnetic structure in \glio projected
onto the orthorhombic ${\bm a}_o{\bm c}_o$ plane [from
Ref.~\onlinecite{gamma}]. b) For comparison, the magnetic
structure in {\em one} honeycomb layer of \alio is plotted for a
scaled propagation vector, $f{\bm q}$, with $f=0.89$, so that it
shows the same periodicity of moment rotation as in \glio. For
both structures left curly arrows indicate counter-rotation of
moments between consecutive sites along ${\bm c}_o$ and both
magnetic structures are plotted for 3 orthorhombic cells along the
horizontal direction. In a) light/dark shaded elliptical envelopes
in unit cell 2 illustrate the alternation of the orientation of
the plane of rotation between adjacent vertically-stacked zigzag
chains, whereas no such alternation occurs in the $\alpha$
magnetic structure (panel b). In unit cell 3 the color of bonds
indicates the anisotropy axis of Kitaev exchange with
black/green/red for $\mathsf{x},\mathsf{y},\mathsf{z}$. The Kitaev
axes are normal to the Ir-O$_2$-Ir bond planes and are shown by
the unit vectors $\bm{\hat{\mathsf{x}}}$, $\bm{\hat{\mathsf{y}}}$
($\bm{\hat{\mathsf{z}}}$ into the page), as defined in
eq.~(\ref{eq:xyz_kitaev}). In b) the axes labels ${\bm a}_m$,
${\bm b}_m$ and ${\bm c}_m$ indicate the ``symmetrized''
monoclinic axes defined in eq.~(\ref{eq:o2m}).}
\label{fig:magstruct_alphagamma}
\end{figure}
%%%%%%%%%%%%%%%%%%%%%%%%%%%%%%%%%%%%%%%%%%%%%%%%%%%%%%%%%%%%%%%%%%%

%% Discussion
\section{Discussion}
\label{sec:discussion} Here we discuss the key features of the
magnetic structure and possible spin Hamiltonians that could
explain its stability. Incommensurate magnetic orders on the
honeycomb lattice have been discussed theoretically for various
frustrated spin
Hamiltonians,\cite{rastelli,rau,rachel,vandenBrink,Chaloupka15}
however the observed counter-rotation of magnetic moments on every
nearest-neighbor bond is a highly non-trivial feature to reproduce
theoretically. As explained in
Refs.~\onlinecite{gamma,universality} for a pair of spins that
counter-rotate the conventional Heisenberg exchange energy is
exactly zero at the mean-field level, i.e. if magnetic moments
${\bm S}_i$ and ${\bm S}_j$ in the unit cell are counter-rotating
in a common plane then $\langle J_{ij}{\bm S}_i \cdot {\bm S}_j
\rangle=0$, where $\langle \ldots \rangle$ means the average over
that type of bond for all unit cells in the crystal. So a spin
Hamiltonian based on dominant Heisenberg exchanges cannot explain
the observed structure. In particular, a Heisenberg model with
couplings up to 3rd nearest neighbor can accommodate
incommensurate moment-rotating ground states in the phase diagram
with propagation vectors along the $a$-axis (the so-called $H1$
and $H3$ phases) or the $b$-axis ($H2$ phase),\cite{rastelli}
however all those phases share the key feature that magnetic
moments are co-rotating, in stark contrast to the
experimentally-observed magnetic structure. Similarly, The
so-called IC$_x$ phase proposed for a model based on frustrated
triplet ferromagnetic dimers\cite{vandenBrink} of \alio does have
Bragg peaks with the same selection rules as plotted in
Fig.~\ref{fig:bz}b) (blue stars), however the ordered moments are
co-rotating, so can also be ruled out. The positions of the
magnetic Bragg peaks can rule out other magnetic structure models,
such as a ``vertex phase"\cite{Chaloupka15} with Bragg peaks at
the corners of the 1st Brillouin zone (inner hexagon in
Fig.~\ref{fig:bz}b), or an incommensurate magnetic
phase\cite{Chaloupka15} continuously connected to the zigzag phase
observed in Na$_2$IrO$_3$; such an incommensurate structure would
have the propagation vector oriented perpendicular to the zigzag
chains (of type ($0,k)$, $k<1$) in the diagram in
Fig.~\ref{fig:bz}b), contrary to the observed wavevector $(q,0)$,
oriented parallel to the zigzag chains.

In order to discuss other spin Hamiltonians that could explain the
stability of the observed magnetic structure in \alio it is
insightful to make a comparison with the magnetic order observed
in the $\beta$ and $\gamma$ polytypes of Li$_2$IrO$_3$, as in all
three cases the magnetic order is incommensurate with magnetic
moments counter-rotating between nearest-neighbor sites. As
explained in Ref.~\onlinecite{universality} the crystal structures
of all three polytypes can be described with reference to a common
orthorhombic cell, which coincides with the structural cell for
the $\beta$ and $\gamma$ structures. In this description the Ir
honeycomb of the $\alpha$ structure is contained in the diagonal
orthorhombic plane ($\bm{a}_o+\bm{b}_o,\bm{c}_o$), where the
subscript $o$ indicates orthorhombic axes. In
Fig.~\ref{fig:magstruct_alphagamma}a-b) we compare the magnetic
structures of the $\gamma$ and $\alpha$ polytypes by looking at
their projection onto the orthorhombic ${\bm a}_o{\bm c}_o$ plane,
this is a convenient comparison as the two iridium lattices are
identical in this projection (see Appendix \ref{app:structure} for
details). For \alio\ (panel b) a hypothetical magnetic structure
is plotted with the same magnetic eigenvector as found
experimentally and plotted in Fig.~\ref{fig:magstruct}, but for a
scaled propagation vector $f{\bm q}$, where $f \simeq 0.89$,
chosen such as to have the same periodicity of the magnetic order
as in the $\gamma$ structure (panel a) (for the scaled propagation
vector $f{\bm q}$ the magnetic moment orientation repeats almost
every 7 zigzag bonds as opposed to a near 6 bonds repeat for the
actual $\alpha$ magnetic structure). Direct comparison between the
two panels of Fig.~\ref{fig:magstruct_alphagamma} shows that,
apart from small variations in the moment amplitudes, the two
magnetic structures are essentially the same up to a single
qualitative difference, which is the fact that the plane of moment
rotation is alternating between vertically-stacked zigzag chains
in the $\gamma$-phase (see light/dark shading of the elliptical
envelopes in unit cell 2), whereas it is not alternating (it is
the same for all zigzag chains) in the $\alpha$ phase. In other
words, the moment components along the Kitaev $\mathsf{z}$ axis
(orthorhombic $b_o$ axis) are ferromagnetically aligned for every
vertical bond in the $\gamma$ (and $\beta$) structures, whereas
they are antiferromagnetically aligned for every vertical bond in
the $\alpha$ structure. These two scenarios are clearly
differentiated by our experiment. An alternation of the rotation
plane would give rise to strong diffraction intensities at
satellite positions of structurally-forbidden reflections, such as
(106)$\pm{\bm q}$, which were not observed (see
Fig.~\ref{fig:hscan}).

It has been theoretically proposed that the $\beta$ and $\gamma$
magnetic structures are stabilized by a dominant Kitaev exchange
supplemented by additional smaller exchange
terms,\cite{gamma,universality,ybkim1,ybkim2} with the Kitaev term
being crucial in stabilizing the counter-rotation of moments. The
strong similarity of this magnetic structure with the one observed
in the $\alpha$ phase, suggests that Kitaev interactions are also
responsible for the counter-rotation of moments in the latter
case. In the $\beta$ and $\gamma$ structures it is understood that
the reason for a tilt of the rotation plane away from the
$\bm{a}_o\bm{c}_o$ plane is the presence of a finite Kitaev
interaction along the vertical ($\mathsf{z}$) bonds
$K_{\mathsf{z}}<0$ (ferromagnetic), which favors an alternating
tilt of the plane of rotation between adjacent zigzag
chains.\cite{universality} The $\alpha$ magnetic structure also
has the plane of rotation tilted away from the $\bm{a}_o\bm{c}_o$
plane, but there is no alternation between adjacent zigzag chains.

In Appendix \ref{app:theory} we perform a soft-spin
analysis\cite{universality} of the magnetic ground state of
candidate spin Hamiltonians that could be compatible with the
observed magnetic structure in \alio. We start with a minimal
nearest-neighbor model that can explain the stability of the
magnetic structures in both $\beta$ and $\gamma$ phases, with
dominant Kitaev interactions $K_{\mathsf z}$ along the vertical
bonds and $K_{\mathsf{x},\mathsf{y}}$ along the zigzag bonds (all
ferromagnetic), an additional smaller (antiferromagnetic)
Heisenberg exchange $J$ on all nearest-neighbor bonds, and an
Ising (ferromagnetic) coupling $I_c$ on the vertical bonds for the
spin components along the bond direction. We find two distinct
modifications of the above Hamiltonian that could explain the
observed eigenvector and pattern of the magnetic structure in
\alio. The first modification, Model A, has the Kitaev interaction
along the vertical bonds having an opposite sign ($K_{\mathsf
z}>0$, antiferromagnetic). The second modification, Model B, has
uniform Kitaev interactions, but is supplemented by an additional
(ferromagnetic) interaction $I_d$ on the zigzag bonds, with
$|I_d|<|I_c|$. In both cases, dominant magnitude Kitaev terms are
required to stabilize the counterrotation of moments.

%% Conclusions
\vspace{1cm}
\section{Conclusions}
\label{sec:conclusions} To summarize, combining single-crystal
magnetic resonant x-ray diffraction and magnetic powder neutron
diffraction on the layered honeycomb $\alpha$-Li$_2$IrO$_3$ we
have observed an incommensurate magnetic structure with
counter-rotating moments for every nearest-neighbor pair of sites.
We have discussed that the counter-rotation of moments cannot be
explained by a spin Hamiltonian with dominant Heisenberg exchange
interactions, and we have compared the observed magnetic structure
with the incommensurate magnetic orders in the three-dimensional
structural polytypes $\beta$- and $\gamma$-Li$_2$IrO$_3$. These
two polytypes also have counter-rotating moments, proposed
theoretically to be stabilized by dominant Kitaev interactions
between spin-orbit entangled $j_{\rm eff}=1/2$ Ir$^{4+}$ magnetic
moments. Based on many striking common features between the
magnetic structures in the three polytypes we have suggested that
Kitaev interactions are the dominant spin couplings that govern
the cooperative magnetism in all three structural polytypes of
Li$_2$IrO$_3$, and using a soft-spin analysis we have proposed a
possible generalization of the spin Hamiltonian used to describe
the $\beta$ and $\gamma$ structures that could account for the
observed magnetic structure in \alio.

% Acknowedgements
\vspace{1cm}
\section{Acknowedgements}
\label{sec:acknowedgements} Work at Oxford was partially supported
by the EPSRC (U.K.) under Grants No. EP/H014934/1 and
EP/M020517/1. RDJ acknowledges support from a Royal Society
University Research Fellowship. IK acknowledges support from an
MIT Pappalardo Fellowship. Work at Augsburg was supported by the
Helmholtz Virtual Institute 521 (``New states of matter and their
excitations") and the German Science Foundation through TRR-80. We
acknowledge Diamond Light Source for time on Beamline I16 under
Proposal MT12028-1.

%%%%%%%%%%%%%%%%%%%%%%%%%%%%%%%%%%%%%%%%%%%%%%%%%%%%%%%%%%%%%%%%%%%%%
\appendix
\section{Magnetic Symmetry Analysis}
\label{app:magnetic}
%%%%%%%%%%%%%%%%%%%%%%%%%%%%%%%%%%%%%%%%%%%%%%%%%%%%%%%%%%%%%%%%%%%%%
Here we give further details of the magnetic symmetry analysis and
the description of the magnetic structure using basis vectors,
following closely the analysis for \blio in
Ref.~\onlinecite{beta}. \alio has a monoclinic crystal structure
with space group\cite{omalley} $C2/m$ and room-temperature lattice
parameters $a=5.1633(2)$~\AA{}, $b=8.9294(3)$~\AA{},
$c=5.1219(2)$~\AA{} and $\beta=109.759(3)^{\circ}$. The iridium
ions occupy a single crystallographic site\cite{nominal} with
multiplicity 4, Wyckoff letter $g$, and site symmetry 2. There are
two iridium atoms per primitive cell, which in the monoclinic
$C$-centered cell correspond to Ir1 at fractional coordinates
($0,y,0$), and Ir2 at ($0,-y,0$), with $y=0.3332$. For the
magnetic propagation vector ${\bm q}=(q,0,0)$ symmetry analysis
using BasiReps\cite{basireps} gives two types of magnetic basis
vectors, $F$ and $A$, which correspond to the case where the
Fourier components, $\bm{M}_{\bm{q},n}$, of the magnetic moments
of the two iridium sublattices ($n=1,2$ for Ir1,2 respectively)
are in phase or in anti-phase, i.e.
$\bm{M}_{\bm{q},2}=\pm\bm{M}_{\bm{q},1}$ with the upper/lower sign
for $F/A$. The irreducible representations of the magnetic
structure and basis vectors are listed in Table~\ref{tab:irr},
where ($x,y,z$) form a Cartesian set of axes related to the
monoclinic axes as $\bm{x} \parallel \bm{a}$, $\bm{y}
\parallel \bm{b}$ and $\bm{z} \parallel \bm{c}^*$.
The magnetic moments are expressed in terms of the Fourier
components as
$\bm{M}_{\bm{r},n}=\sum_{\bm{k}=\pm\bm{q}}\bm{M}_{\bm{k},n}e^{-i\bm{k}\cdot\bm{r}}$,
where $\bm{M}_{-\bm{q},n}=\bm{M}^*_{\bm{q},n}$ as the magnetic
moment distribution is real.

The requirement that the magnetic structure is invariant under
symmetry operations of the full group that maps $\bm{q}$ into
$-\bm{q}$ (in the preset case a two-fold axis $2_y$ at the iridium
sites) imposes additional constraints onto the relative phases
between basis vector components. For an incommensurate propagation
vector perpendicular to the two-fold axis the symmetry-allowed
magnetic structures can be of the following two types: i)
collinear, amplitude-modulated (spin-density-wave type) with
magnetic moment either along the $y$-axis or in a general
direction in the perpendicular $xz$ plane, or ii) moment-rotating,
with an elliptical envelope with a principal axis along $y$ (no
other magnetic structures remain invariant under the $2_y$
rotation). In particular, for basis vectors belonging to the
$\Gamma_1$ irreducible representation the allowed phase
combinations (verified using the ISODISTORT\cite{isodistort}
software) are ($\pm i A_x,F_y,\pm i A_z$) with unconstrained
magnetic moment magnitudes $M_x$, $M_y$ and $M_z$. The
experimentally-determined magnetic structure ($-iA_x,F_y,-iA_z$)
is indeed one of those combinations. In this case the Fourier
components of the magnetic structure are
\begin{equation} \bm{M}_{\bm{q},n} =   \mp i \left(
\bm{\hat{x}} \frac{M_x}{2} + \bm{\hat{z}} \frac{M_z}{2} \right) +
\bm{\hat{y}} \frac{M_y}{2},
 \label{eq:Mq}
\end{equation}
where the upper (lower) sign is to be used for the $n=1 (2)$
sublattice and $\bm{\hat{x}}$ indicates a unit vector along the
$x$-direction and so on. The magnetic moment at position $\bm{r}$
belonging to site index $n$ is obtained via direct Fourier
transformation from (\ref{eq:Mq}) as
\begin{equation} \bm{M}_{\bm{r},n}  =  \mp \left( \bm{\hat{x}} M_x
 + \bm{\hat{z}} M_z \right) \sin
\bm{q}\cdot\bm{r} + \bm{\hat{y}} M_y \cos \bm{q}\cdot\bm{r}.
\label{eq:Mr}
\end{equation}
The above equation describes all iridium sites, including those
related by $C$-centering translations, where $\bm{r}$ is the
actual position of the ion and $n$ is the site index at the
equivalent position (1,2) in the primitive unit cell. The magnetic
structure is plotted in Fig.~\ref{fig:magstruct}a) and shows
magnetic moments rotating between sites displaced along the
(horizontal) propagation direction, and describing an elliptical
envelope with a principal axis along $b$. The invariance of the
magnetic structure with respect to a $2_y$ rotation is most easily
visualized in the extended plot in
Fig.~\ref{fig:magstruct_alphagamma}b) which displays many more
sites (scaling of the propagation vector keeps the symmetry
properties unchanged): here one can see that the spin order is
invariant upon a two-fold rotation around a vertical axis located
where the spin moment is aligned vertically (the 8th spin along
the horizontal zigzag chain), so the magnetic structure is indeed
compatible with the full symmetry of the space group.

\section{The crystal and magnetic structure of
$\bm{\alpha}$-L\MakeLowercase{i}$_2$I\MakeLowercase{r}O$_3$ in the
orthorhombic basis} \label{app:structure}

As explained in Ref.~\onlinecite{universality} the crystal
structures of all three polytypes of Li$_2$IrO$_3$ could be
described in terms of a common orthorhombic unit cell, with
lattice vectors (subscript $o$) related to the monoclinic axes
vectors by
\begin{eqnarray}
{\bm a}_o&=&{\bm a}_m+{\bm c}_m,\nonumber\\
{\bm b}_o&=&{\bm a}_m-{\bm c}_m,\nonumber\\
{\bm c}_o&=&2{\bm b}_m,\label{eq:o2m}
\end{eqnarray}
where the subscript $m$ indicates a ``symmetrized'' monoclinic
cell where the lattice parameters satisfy $a_m:c_m=1:1$ (in the
actual crystal structure of \alio this ratio is\cite{omalley}
$1.008:1$). The orthorhombic description also has the advantage
that the anisotropy (Kitaev) axes associated with each bond
(direction normal to the Ir-O$_2$-Ir planes), are easily
visualized, in particular\cite{modic}
\begin{eqnarray}
\mathsfbf{\hat{x}} & =
&(\bm{\hat{a}}_o+\bm{\hat{c}}_o)/\sqrt{2},\nonumber\\
\mathsfbf{\hat{y}} &
=&(\bm{\hat{a}}_o-\bm{\hat{c}}_o)/\sqrt{2},\nonumber\\
\mathsfbf{\hat{z}} & = & \bm{\hat{b}}_o,\label{eq:xyz_kitaev}
\end{eqnarray}
where we have assumed an ``idealized" crystal structure with cubic
IrO$_6$ octahedra and lattice parameters in ratio $a_o : b_o : c_o
= 1 : \sqrt{2} : 3$. We use SansSerif symbols for the Kitaev axes
(${\mathsf x}$,${\mathsf y}$,${\mathsf z}$) to distinguish them
form the italic symbols ($x$,$y$,$z$), which denote the Cartesian
axes used to describe the magnetic structure. The Kitaev axes are
shown in unit cell 3 in Fig.~\ref{fig:magstruct_alphagamma}a-b)
where the color of the bonds indicates the anisotropy axis of the
Kitaev exchange.

The iridium lattices in the Li$_2$IrO$_3$ polytypes can be thought
of as being constructed from zigzag chains that run along one of
the two diagonal directions in the orthorhombic basal plane ${\bm
a}_o\pm{\bm b}_o$, connected by vertical bonds along ${\bm c}_o$.
In the $\alpha$ structure all zigzag chains are in the diagonal
plane (${\bm a}_o+{\bm b}_o$,${\bm c}_o$), and are connected
vertically to form a honeycomb lattice. In the $\gamma$ structure
pairs of zigzag chains form coplanar honeycomb strips ($33'$ with
$11'$ and $22'$ with $44'$ in
Fig.~\ref{fig:magstruct_alphagamma}a) that are then stacked along
the vertical direction ${\bm c}_o$ alternating in orientation
between the two diagonal planes (${\bm a}_o + {\bm b}_o$,${\bm
c}_o$) and (${\bm a}_o - {\bm b}_o$,${\bm c}_o$); in the $\beta$
structure single zigzag chains alternate in orientation between
the two diagonal directions. From this description it follows that
the projection onto the ${\bm a}_o{\bm c}_o$ and ${\bm b}_o{\bm
c}_o$ planes is then the same in all three structures (the
projections onto the ${\bm a}_o{\bm b}_o$ plane are different).

In the orthorhombic axes notation, the Fourier components of the
magnetic structure, eq.~(\ref{eq:Mq}), are
\begin{equation} \bm{M}_{\bm{q},n} =   \mp i \left(
\bm{\hat{x}}_o \frac{M_{x_o}}{2} - \bm{\hat{y}}_o
\frac{M_{y_o}}{2} \right) + \bm{\hat{z}}_o \frac{M_{z_o}}{2},
 \label{eq:Mqo}
\end{equation}
where the upper/lower sign is to be used for the sublattices
Ir1/Ir2 (which correspond to sites $1^{\prime}/2^{\prime}$
respectively in the $\gamma$ structure, see
Fig.~\ref{fig:magstruct_alphagamma}) and
$M_{x_o}=M_x\cos(\beta/2)+M_z\sin(\beta/2)$,
$M_{y_o}=-M_x\sin(\beta/2)+M_z\cos(\beta/2)$, $M_{z_o}=M_y$. Here
$\bm{\hat{x}}_o$, $\bm{\hat{y}}_o$, $\bm{\hat{z}}_o$  are unit
vectors along the orthorhombic ${\bm a}_o$, ${\bm b}_o$ and ${\bm
c}_o$ axes. For the determined magnetic structure the moment
magnitudes are $M_{x_o}$:$M_{y_o}$:$M_{z_o}$=$0.67$:$0.33$:$1$ and
the moment rotation plane makes an angle
$\phi=\tan^{-1}(M_{y_o}/M_{x_o})=26^\circ$ with the ${\bm
a}_o\bm{c}_o$ plane. The magnetic moment expression,
eq.~(\ref{eq:Mr}), transforms to
\begin{equation} \bm{M}_{\bm{r},n}  =  \mp \left( \bm{\hat{x}}_o
M_{x_o}
 - \bm{\hat{y}}_o M_{y_o} \right) \sin
\bm{q}\cdot\bm{r} + \bm{\hat{z}}_o M_{z_o} \cos \bm{q}\cdot\bm{r}.
\label{eq:Mro}
\end{equation}

\section{Counterrotation of moments and the interference term
in the scattering cross-section}
\label{app:interference}

Here we present an intuitive explanation of the anti-phase
behavior of the magnetic scattering intensity between the
satellites at $(116)\pm\bm{q}$ illustrated in
Fig.~\ref{fig:azimuths}b-c), namely at azimuth values $\Psi$ where
the $+\bm {q}$ satellite is strong the $-\bm{q}$ satellite is
weak, and vice versa. We will show that this qualitative feature
of the magnetic scattering can only be explained by
counter-rotating moments on the two Ir magnetic sublattices. In
this case the scattering intensity contains an interference term
that changes sign between the two satellite positions, naturally
explaining the observed intensity behavior.

To highlight the main terms in the scattering cross-section we
first assume $M_x=0$, i.e. we neglect the contribution from the
small moment components along the $x$-direction. In this case the
total magnetic structure factor vector for a magnetic satellite at
$\bm{Q}=(h,k,l)\pm\bm{q}$ is
\begin{equation}
\bm{{\cal{F}}}(\bm{Q})=\bm{\hat{y}}{\cal{S}}_y M_{y,\pm \bm{q},1}
+ \bm{\hat{z}} {\cal{S}}_z M_{z,\pm \bm{q},1},\label{eq:f}
\end{equation}
where ${\cal{S}}_{y/z}$ are the structure factors of the magnetic
basis vectors along the $y/z$ directions, $M_{y/z,\pm \bm{q},1}$
are the corresponding Fourier components of the magnetic moments
on the Ir1 sublattice and $(x,y,z)$ are Cartesian axes fixed with
respect to the crystal axes, as defined previously. Without loss
of generality we take $M_{y,\pm \bm{q},1}=M_y/2$ and $M_{z,\pm
\bm{q},1}=e^{\pm i\varphi}M_z/2$, where $M_{y/z}$ are the (real)
magnetic moment magnitudes along the $y/z$ axes and $\varphi$ is
the relative phase between the $y$ and $z$ components, constrained
by symmetry to be an integer multiple of $\pi/2$ (see Appendix
\ref{app:magnetic}).

In the experimental scattering geometry employed, as indicated in
Fig.~\ref{fig:azimuths}(inset), the magnetic Bragg peak intensity
depends only on the projection of the total magnetic structure
factor vector $\bm{{\cal{F}}}$ onto the scattered wavevector
direction $\bm{\hat{k^\prime}}$, see Ref.~\onlinecite{hill}. In
detail, the intensity is proportional to
\begin{eqnarray}
\left| \bm{{\cal{F}}}\cdot{\bm{\hat{k^{\prime}}}}\right|^2 & = [
\left|\hat{k'_y} M_y {\cal{S}}_y\right|^2+
\left|\hat{k'_z} M_z {\cal{S}}_z\right|^2  \nonumber \\
& + ~ 2 \hat{k'_y} \hat{k'_z} M_y M_z {\cal{A}} ]/4,
\label{eq:intensity}
\end{eqnarray}
where the first two terms are the separate magnetic scattering
intensities from the $y$ and $z$ moments, and the last term is due
to interference scattering between the $y$ and $z$ moments. The
intensity dependence on the azimuth comes exclusively from the
projections $\hat{k'_y}$ and $\hat{k'_z}$ of $\bm{\hat{k^\prime}}$
onto the $y$ and $z$ directions, respectively. The interference
term in eq.~(\ref{eq:intensity}) is directly sensitive to the
basis vector combination through the factor
\begin{equation}
{\cal{A}}={\cal{R}}\cos\varphi \pm {\cal{I}}\sin \varphi,
\label{eq:interference}
\end{equation}
where ${\cal{R}}$ and ${\cal{I}}$ are the real and imaginary parts
of the product ${\cal{S}}_y{\cal{S}}^*_z$ and the upper/lower sign
corresponds to the $\pm{\bm q}$ satellite. If the interference
factor ${\cal{A}}$ cancels, then the azimuth dependence of the
intensity in eq.~(\ref{eq:intensity}) is essentially the same
between paired satellites (up to variations in the geometrical
factors between the two satellites, which are expected to be small
if the two wavevectors are close, i.e. if $|(hkl)| \gg |\bm{q}|$).
To obtain a large intensity difference between paired satellites
the interference term needs to be large and to alternate in sign
between the two satellites, i.e. ${\cal{I}}\sin\varphi\neq 0$ in
eq.~(\ref{eq:interference}). Below we analyze all possible
combinations of basis vectors and relative phases and find that a
sign-alternating interference term occurs only if the basis
vectors along the two directions are of different type and are
$\pi/2$ out of phase, which corresponds to a magnetic structures
with counter-rotating moments on the two sublattices. This follows
from the fact that the magnetic structure factors for the two
basis vectors in eq.~(\ref{eq:sf}) are either purely real ($F$) or
purely imaginary ($A$), so the product ${\cal{S}}_y{\cal{S}}^*_z$
is purely real (${\cal{R}}\neq 0$ and ${\cal{I}}=0$) if the basis
vectors along the two directions are the same, or purely imaginary
(${\cal{R}}=0$ and ${\cal{I}}\neq 0$) if they are different. Based
on this observation we identify four distinct cases summarized in
Table~\ref{tab:interference}:

%%%%%%%%%%%%%%%%%%%%%%%%%%%%%%%%%%%%%%%%%%%%%%%%%%%%%%%%%%%%%%%%%%%
\begin{table}[!tbp] \caption{\label{tab:interference}
Properties of the interference factor ${\cal A}$ in the magnetic
scattering intensity in eq.~(\ref{eq:intensity}) at
$(hkl)\pm{\bm{q}}$ as a function of the magnetic basis vector
combination in the ground state.}
\par
\begin{center}
\begin{tabular}{c|c|c|c}
Basis    & Phase     & Magnetic  &  Interference    \\
Vectors  & $\varphi$ & Structure &  factor ${\cal{A}}$          \\
\hline
$FF$  & $0,\pi$ & collinear &  ${\cal{R}} \cos \varphi$ \\
$AA$  &         & SDW       &   \\
\hline
$FA$  & $0,\pi$ & non-collinear &  $0$\\
         &         & SDW           &   \\
\hline
$FF$  & $\pi/2,-\pi/2$ & co-rotating &  $0$\\
$AA$  &                &             &   \\
\hline
$FA$  & $\pi/2,-\pi/2$ & counter-rotating &  $\pm{\cal{I}}\sin\varphi$\\
\end{tabular}
\end{center}
\end{table}
%%%%%%%%%%%%%%%%%%%%%%%%%%%%%%%%%%%%%%%%%%%%%%%%%%%%%%%%%%%%%%%%%%%

i) same basis vectors along the two directions with relative phase
$\varphi=0$ or $\pi$, the magnetic structure is a
spin-density-wave (SDW), collinear between the two sublattices,
the interference term factor is finite and has the same sign for
paired satellites,

ii) different type basis vectors with relative phase $\varphi=0$
or $\pi$, each sublattice has a spin-density-wave order, but
non-collinear between the two sublattices, there is no
interference term

iii) same basis vectors on the two directions with relative phase
$\varphi=\pi/2$ or $-\pi/2$, magnetic moments co-rotate on the two
sublattices (the sign of $\varphi$ gives the absolute sense of
rotation on the Ir1 sublattice), here also there is no
interference term, and

iv) different basis vectors with relative phase $\varphi=\pi/2$ or
$-\pi/2$, magnetic moments counter-rotate on the two sublattices,
the interference term is finite, sign-alternating between paired
satellites and with the absolute sign determined by whether the
rotation at site Ir1 is clockwise or counterclockwise.

The observation of an anti-phase behavior of the intensity between
the paired satellites at $(116)\pm{\bm{q}}$ [see
Fig.~\ref{fig:azimuths}b-c)] can only be explained by a magnetic
structure of type iv) above, with counter-rotating moments. The
effect can be understood in terms of a sign-alternating
interference scattering term between the $y$ and $z$ magnetic
moment components, coupled in basis vectors of $F$ and $A$ type,
respectively, with a relative phase $\varphi=-\pi/2$ (in the main
text we have used the shorthand notation ($F_y,e^{i\varphi}A_z$)
to denote this basis vector combination). Since the wavevector
$\bm{Q}$ is close to the $z$-axis the azimuth dependence of the
geometrical factors are (to leading order) $\hat{k'_z}\simeq\sin
\theta$ and $\hat{k'_y}\simeq -\cos \theta \sin \Psi$, where
$2\theta$ is the total scattering angle; so the largest magnitude
interference term in eq.~(\ref{eq:intensity}) (largest contrast
between the intensities of the paired satellites) is expected for
$\Psi$ near $\pm90^\circ$, as indeed observed. The red solid lines
in Fig.~\ref{fig:azimuths}b-c) show the calculated intensity
including the full azimuth dependence of the geometrical factors
and also the effect of the small, but finite $M_x$ moment
components.

\section{Minimal spin Hamiltonian for
$\bm{\alpha}$-L\MakeLowercase{i}$_2$I\MakeLowercase{r}O$_3$}
\label{app:theory}

Here we study a range of minimal spin Hamiltonians, based on
nearest-neighbor exchanges only, seeking to capture the observed
incommensurate magnetic structure. The important features can be
qualitatively summarized as follows: (1) counter-rotation between
the Ir1/2 sublattice moments, and (2) uniformly tilted plane of
rotation. Here by ``plane of rotation'' we refer to the plane in
the Bloch sphere spanned by the various magnetic moments across
the lattice; using the orthorhombic axes notation as in
eq.~(\ref{eq:Mro}), this is the plane spanned by the vectors
$(\bm{\hat {x}}_o M_{x_o} - {\bm{\hat{y}}}_o M_{y_o})$ and
${\bm{\hat{z}}}_o $. Feature (1) has been previously
shown\cite{universality} to require anisotropic (i.e.
non-Heisenberg) nearest-neighbor exchange and in particular is
consistent with strong Kitaev exchange. In a minimal
nearest-neighbor model feature (1) can be
captured\cite{universality} by (ferromagnetic) Kitaev exchanges on
all nearest neighbor bonds $K_{\mathsf z}$,
$K_{\mathsf{x},\mathsf{y}}<0$, a smaller (antiferromagnetic)
Heisenberg exchange $J>0$ on all those bonds and a finite
(ferromagnetic) exchange $I_c<0$ for the vertical bonds, which
couples the moment components along the bond direction. Here we
discuss modifications that are appropriate for feature (2).

%%%%%%%%%%%%%%%%%%%%%%%%%%%%%%%%%%%%%%%%%%%%%%%%%%%%%%%%%%%%%%%%%%%
\begin{figure}[!tbh]
\includegraphics[width=\columnwidth]{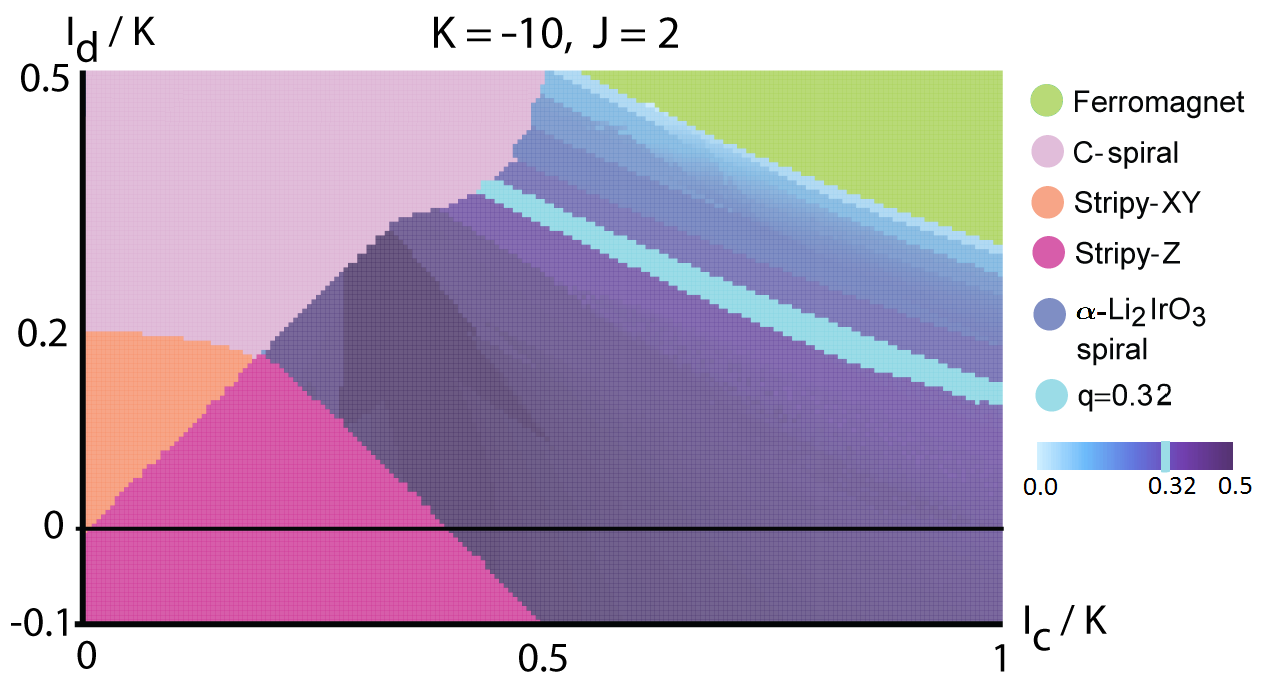}
\caption[]{(color online) Phase diagram of Model B described in
the text, computed in a soft spin approximation. We find that the
observed counter-rotating magnetic order ($q=0.32$, light blue
line), with a uniformly tilted plane of rotation, can be captured
by adding ``truncated-dipole'' superexchange interactions, $I_d$
and $I_c$, within the regime $|I_c|>|I_d|$,  $I_d/K>0$. These
interactions couple spins to the spatial honeycomb plane, which
combines with the counter-rotation due to Kitaev exchange to
produce the experimentally observed magnetic order pattern. The
observed spiral phase is shown in a color gradient corresponding
to the magnitude of the propagation vector $q$, from 0 to $0.5$ in
units of $2\pi/a$. Comparing to Fig.~\ref{fig:bz}b), the
wavevector is along the horizontal direction, with units such that
$q=1$ would correspond to the white diamond symbol. Labels
stripy-XY and stripy-Z denote antiferromagnetic patterns where
spins are aligned with one of their three nearest-neighbors,
across a diagonal/vertical bond for stripy-XY/Z respectively, and
antialigned with the other two. Label C-spiral denotes an
incommensurate counterrotating order with propagation vector along
the vertical ($\bm{\hat{c}}_o=\bm{\hat{b}}_m$) direction in
Figs.~\ref{fig:bz}a,b).} \label{fig:phasediagram}
\end{figure}
%%%%%%%%%%%%%%%%%%%%%%%%%%%%%%%%%%%%%%%%%%%%%%%%%%%%%%%%%%%%%%%%%%%

We have found two independent modifications that can produce
feature (2) in a dominant-Kitaev Hamiltonian. In general, both
modifications could occur and could complement each other.  Here
we consider them separately. The first modification, denoted as
Model A, consists of a sign change for the Kitaev exchange on
``vertical'' bonds, i.e. those bonds which lie parallel to the
crystalline $\bm{c}_o$ (or equivalently $\bm{b}_m$) axis, i.e.
$K_{z}>0$. In this model, the $\bm{b}_o$-axis moment components
would be anti-aligned along those bonds, straightforwardly
producing the observed tilt pattern.

The second modification, denoted as Model B, consists of an
additional weaker exchange on the zigzag bonds, i.e. the
nearest-neighbor bonds that are not parallel to the crystalline
$\bm{c}_o$ axis. This exchange, which we denote by $I_d$, is a
symmetry-allowed nearest-neighbor exchange, which couples the spin
components pointing along the bond direction
\begin{align}
I_d S_i^r S_j^r ~,~ \quad S^r \equiv \bm{S} \cdot {\bm{\hat{r}}}
\end{align}
where $\bm{r}$ is the vector connecting sites $i$ and $j$. This
exchange interaction has a form that is mathematically analogous
to a truncated dipole interaction (though its physical origin is
through superexchange). It is related to the $\Gamma$ spin
exchange term that has been previously discussed in the context of
the layered honeycomb iridates.\cite{rau,universality} We use the
subscripts on $I_c$ and $I_d$ to denote that the value of the
exchange can differ between the ``vertical'' (along $\bm{c}_o$)
and the remaining (zigzag or ``diagonal'') bonds.

Here we start with a model with dominant ferromagnetic Kitaev
exchanges, which for simplicity we take to have the same magnitude
(and sign) on all three nearest-neighbor bonds ($K_{\mathsf
z}=K_{{\mathsf x},{\mathsf y}}=K<0$), and an additional small
antiferromagnetic Heisenberg exchange $J$ on all those bonds. We
find that adding ferromagnetic $I_c$ and $I_d$ exchanges (i.e. of
the same sign as the Kitaev exchange) can stabilize the observed
counterrotating magnetic order with a uniformly tilted plane of
rotation, if $I_d/K>0$ and $|I_c|>|I_d|$.

A representative soft-spin phase diagram is shown in
Fig.~\ref{fig:phasediagram}. The observed \alio counter-rotating
order is seen across a range of parameters, with a propagation
vector that varies continuously across the parameter space, and
which includes the experimentally observed value (light blue
shading). The lowest energy mode in this phase is seen to have a
plane of rotation that is tilted uniformly. The sign of the tilt
agrees with the sign observed experimentally, namely it is a small
tilt, away from the $\bm{a}_o\bm{c}_o$ plane, in the direction
away from the plane of the honeycomb lattice.

Within the soft spin approximation, the magnitude of the tilt
angle is seen to be about half of the experimentally observed
value for typical parameter points with the observed wavevector,
and in general varies across the phase as the $I_d$ interactions
are turned on. However, we note that the soft spin ground state
here requires the spins to be soft and have non-uniform magnitude,
representing strong thermal or quantum fluctuations, and thus it
is not expected to capture the tilt angle quantitatively. For
instance, a sample set of Hamiltonian parameters can be chosen as
\begin{align} K=-10, \ J=2, \ I_c=-5, \ I_d = -3.5 \end{align}
(where the energy unit may be taken as $\sim$0.45~meV in order to
match the observed $T_{\rm N}$). For those parameters the
propagation vector agrees with the experimental value $q=0.32(1)$
and the ordered spin magnitudes, using the orthorhombic axes as
per eq.~(\ref{eq:Mro}), are found to be in the ratio
\begin{align} S_{x_o}:S_{y_o}:S_{z_o} = 0.56:0.15:1. \end{align}

The qualitative feature of uniform tilt is captured by this
analysis, and may be understood as resulting from the uniform
spatial plane of the honeycomb lattice: the $I_c$ and $I_d$
exchanges couple the spins to the orientation of the bonds on the
honeycomb lattice, and can thereby produce this uniform tilt. The
sign of the tilt, which is set to be away from the honeycomb
plane, is produced by the counter-rotation of adjacent
sublattices, which sets the $\bm{b}_o$ components of spins to be
anti-aligned between neighboring sites (as per
eq.~(\ref{eq:Mro})), and is thus tilted away from the spatial
honeycomb plane by a small ferromagnetic $I_d$ exchange.


\begin{thebibliography}{99}
\bibitem{error} The uncertainty in the value of $q$ is estimated
from combining uncertainties in the crystal orientation
(UB-matrix) and the I16 instrument intrinsic sphere of
uncertainty.
\bibitem{nominal}
We use the nominal crystal structure with no site mixing, where Ir
ions occupy the single crystallographic site $4g (0,y,0)$ in the
$C2/m$ cell of Ref.~\onlinecite{omalley}.
\end{thebibliography}
\end{document}